\documentclass[lettersize,journal]{IEEEtran}
\IEEEoverridecommandlockouts

\usepackage{cite}
\usepackage{amsmath,amssymb,amsfonts}
\usepackage{algorithmic}
\usepackage{graphicx}
\usepackage{textcomp}
\usepackage{xcolor}
\usepackage{diagbox}
\usepackage{mathtools}
\usepackage{float}
\usepackage{tabularx}
\usepackage{booktabs}
\usepackage{caption} 
\usepackage{subfigure} 

\usepackage[bookmarks=false]{hyperref}
\usepackage[letterpaper, top=0.7in, bottom=1in, left=0.625in, right=0.625in]{geometry}
\usepackage{amsthm}
\newtheorem{theorem}{Theorem}

\usepackage{multirow} 
\usepackage{makecell}
\usepackage{hyperref} 

\usepackage[ruled,linesnumbered]{algorithm2e}

\DeclareMathOperator{\diag}{diag}

\DeclareMathOperator{\Var}{Var}

\usepackage{bm}

\newcommand{\norm}[1]{\left\lVert #1 \right\rVert}

\newcommand{\set}[1]{\left\{ #1 \right\}}

\begin{document}

\title{Uncertainty-Aware Distributed Radio Map Estimation in Decentralized Wireless Networks
\thanks{This work was supported in part by the NSF grants 2146497, 2231209, 2244219, 2315596, 2416872, 2533587, 2548961, 2551418, and 2551417.}
}

\author{\IEEEauthorblockN{Dong Yang$^\dagger$, \;Zhuoyu Yao$^\dagger$, \; Yue Wang$^\dagger$, \; Songyang Zhang$^*$, \; Yingshu Li$^\dagger$, \; Zhi Tian$^\ddagger$, \; Zhipeng Cai$^\dagger$}
\IEEEauthorblockA{$^\dagger$Department of Computer Science, Georgia State University, Atlanta, GA, USA\\$^*$Department of Electrical and Computer Engineering, University of Louisiana at Lafayette, LA, USA\\$^\ddagger$Department of Electrical and Computer Engineering, George Mason University, Fairfax, VA, USA}}

\maketitle
\thispagestyle{empty}

\begin{abstract}
Federated learning based radio map estimation (RME) approaches rely on frequent exchange of high-dimensional 
model updates and often require a central server for collaboration coordination and update aggregation. 
This incurs prohibitively high communication overhead, rendering them impractical for bandwidth-constrained mobile and wireless edge networks. 
Moreover, in practical mobile sensing scenarios, sparse and unevenly collected measurements lead to estimation uncertainty, which is critical for guiding RME and collaboration among distributed nodes. 
To address these challenges, this paper proposes an uncertainty-aware distributed radio map estimation (UA-D-RME) framework for decentralized networks. 
In UA-D-RME, 
each 
node 
locally infers a probabilistic model from its sparse measurements 
to estimate both its local radio map and the associated estimation uncertainty.
Instead of exchanging raw data or model updates, neighboring nodes communicate only low-dimensional  statistical information, 
including radio map and uncertainty measures over their overlapping monitoring areas.
%
An uncertainty-aware statistical consensus mechanism is then developed to iteratively align  probabilistic models in neighborhood by ensuring consistency on estimation confidence. 
Theoretical analysis establishes the convergence of the proposed consensus mechanism and characterizes the distributed generalization performance of the proposed UA-D-RME method.
Experiments on urban radio map datasets verify that UA-D-RME achieves desired reconstruction accuracy comparable to the centralized RME benchmark while 
reducing communication costs and without exchanging raw measurements.
\end{abstract}


\section{Introduction}
\label{sec:intro}
The evolution of wireless 
systems 
has created an unprecedentedly dense and dynamic radio frequency (RF) environment. Beyond traditional spectrum sensing in cognitive radios\cite{wang2012sparsity}, radio map estimation (RME) is a fundamental enabler to infer the spatial distribution of critical radio quantities\cite{11677060}, such as received power, interference levels, and power spectral density (PSD), from sparse and geographically distributed measurements~\cite{yang2024spectrum}. 
High-fidelity radio maps are indispensable for a wide range of advanced applications, including network planning, low-altitude networks, and the construction of wireless digital twins. 
As wireless systems evolve toward dense, heterogeneous, and edge-native architectures, reconstructing accurate radio maps from limited measurements becomes increasingly critical. 

Traditional model-based methods, such as leveraging log-distance path loss~\cite{lee2012voronoi} and inverse distance weighting~\cite{kuo2010discriminant}, 
rely on simplified propagation assumptions or interpolation techniques.
However, these methods 
often degrade in realistic urban environments dominated by complex blockage, shadowing, and local scattering. 
Alternatively, data-driven 
RME solutions have emerged to learn complex propagation patterns from measurement data.
Representative studies include convolutional radio-map completion \cite{levie2021radiounet}, 
deep autoencoders \cite{teganya2021deep},  tensor-based spectrum cartography \cite{shrestha2022deep}, 
graph neural approaches~\cite{liu2026collaborative}, and transformer-based estimators \cite{nan2026cgformer}. 
Recently, advanced generative architectures, such as diffusion-based Bayesian RME~\cite{ha2025bayesian}, and foundation-model priors \cite{yang2026fm}, have further improved the ability of data-driven models to capture underlying RF features.


Despite such remarkable advances, existing RME methods are usually built upon a centralized learning paradigm, in which raw measurement data collected by distributed sensing nodes must be uploaded to a central server. 
Such centralized data collection incurs substantial communication overhead and raises privacy concerns, especially in communication-constrained mobile systems and wireless edge networks\cite{Wang2024Distributed}.  
To mitigate these issues, federated learning (FL) has been introduced to distributed RME, enabling multiple nodes to collaboratively train a model while keeping their raw data local~\cite{yang2025physics}. 
However, standard FL still 
relies on the frequent exchange of high-dimensional model parameters or gradients, which makes FL unsuitable for mobile and wireless networks with limited link bandwidth and  energy budget\cite{zhou2025efficient}. 
Furthermore, standard FL typically depends on 
a central parameter server for collaboration control and globally coordinated model update aggregation, which prevents FL applicability 
in ad-hoc topology of edge networks~\cite{yao2026generative}. 


Another critical yet underexplored issue is inherent estimation uncertainty in distributed RME literature\cite{11677060}. 
In practical sensing scenarios, measurements are often sparse, unevenly distributed, and locally biased due to mobility constraints, sensing costs, and heterogeneous propagation environments. 
As a result, distributed nodes inevitably exhibit different levels of estimation uncertainty across space. 
Ignoring such uncertainty information, distributed nodes may treat unreliable estimates and confident estimates equally during collaboration, degrading the performance of distributed RME. 
In this sense, uncertainty quantification is essential not only for assessing the reliability of 
RME, but also for guiding collaboration 
among nodes.

To address the above challenges, we propose an uncertainty-aware distributed RME (UA-D-RME) framework for decentralized networks. 
We first formulate RME as a decentralized collaborative learning problem. 
In UA-D-RME, each node locally infers a probabilistic model from its sparse measurements to estimate both its local radio map and the associated estimation uncertainty.
Instead of transmitting raw data or high-dimensional model updates, neighboring nodes exchange only low-dimensional predictive statistics, including on-demand estimated radio map values and uncertainty levels, over their overlapping monitoring areas.
Based on the exchanged statistics, an uncertainty-aware statistical consensus mechanism is developed to iteratively align neighboring probabilistic models. 
By weighting neighborhood consistency per uncertainty, UA-D-RME enables distributed nodes to maintain calibrated uncertainty estimates and improve RME accuracy under communication constraints. 
Our contributions 
are listed below.
\begin{enumerate}
\item {Decentralized RME:}  To the best of our knowledge, this is the first work of distributed RME in decentralized network topology. We formulate 
the communication-constrained RME problem in a decentralized manner. 
This formulation shifts  RME paradigm from  conventional centralized solutions toward decentralized, region-wise, autonomous 
 coordination for efficient distributed RME. 
\item {Uncertainty-Aware Statistics Consensus:} We propose a decentralized RME algorithm empowered by a variational autoencoder, where each node maintains a local probabilistic model. 
Based on consensus optimization, 
each node 
iteratively reaches estimation consistency with neighbors at lower 
communication cost than benchmark methods, by exchanging only uncertainty-aware radio statistics across the shared monitoring areas.
\item {Theoretical Analysis:} We provide  theoretical insights into the 
UA-D-RME 
by analyzing the convergence of the uncertainty-aware consensus mechanism and characterizing the distributed generalization performance. It demonstrates that the confidence-weighted collaboration 
improves neighborhood estimation consistency and 
supports reliable and efficient 
decentralized RME.  
\item {Empirical Validation:} 
We conduct experiments on urban radio map datasets to evaluate 
UA-D-RME, 
which achieves desired estimation accuracy comparable to a centralized RME benchmark at the lower communication cost, while 
keeping raw measurements local. 
We open-source the code\footnote{The code is available at {\url{https://github.com/DongYang26/UA-D-RME}}.}, in order to facilitate reproducibility, fair comparison, and future research on distributed RME.
\end{enumerate}

\section{System Model and Problem Formulation}
\label{sec:system}

\subsection{Spatial Distribution of Received PSD}
We first 
establish a physical model to characterize the received PSD across various spatial locations within a specific target region 
\( \mathcal{X} \subset \mathbb{R}^2 \). 
Consider \( M \) active transmitters 
operating within a designated frequency band in this region of interest. 
Let \( \Gamma_m(f) \) denote the transmit PSD of the \( m \)-th transmitter 
at frequency 
\( f \in \mathcal{F} \), where $\mathcal{F}$ represents the set of operational frequencies. Define \( G_m(\boldsymbol{\mathit{x}}, f) \) as the channel frequency response from the \( m \)-th transmitter to a mobile sensing node equipped with an isotropic antenna at location \( \boldsymbol{\mathit{x}} \in \mathcal{X} \). For a single observation snapshot, 
\( \Gamma_m(f) \) and \( G_m(\boldsymbol{\mathit{x}}, f) \) are time-invariant over a 
short observation period, thereby yielding a stationary ground-truth radio map. 

Assuming that the \( M \) transmitters operate independently, 
the received PSD at 
location \( \boldsymbol{\mathit{x}} \in \mathcal{X} \) at frequency \( f \in \mathcal{F} \) can be formulated as:
\begin{equation}\textstyle
    \label{eq:PSD_y}
    \mathbf{\Phi}(\boldsymbol{\mathit{x}}, f) = \sum_{m = 1}^{M} \Gamma_m(f) |G_m(\boldsymbol{\mathit{x}}, f)|^2 + \nu(\boldsymbol{\mathit{x}} , f),
\end{equation}
where 
\( \nu(\boldsymbol{\mathit{x}} , f) \) denotes 
the additive white Gaussian noise at location \( \boldsymbol{\mathit{x}} \) and frequency \( f \). 
For simplicity but without loss of generality, we hereafter focus on 
the single-frequency case, while the methodology and solutions proposed in this work can be extended to the multiband case as well. 

\subsection{Mobile Edge Nodes and Communication Topology}
Let $\mathcal{A} = \set{1,\ldots,K}$ denote a set of distributed mobile edge nodes serving as local sensing and processing units. 
Each node $i \in \mathcal{A}$ is responsible for monitoring a local region 
$\mathcal{X}_i \subset \mathcal{X}$, and the union of all local regions covers the entire target region 
$\bigcup_{i=1}^{K} \mathcal{X}_i = \mathcal{X}$. 
In practice, it is worth noting that adjacent nodes usually have partially overlapping geographical coverage areas, 
which motivates collaborative RME in neighborhood. 
For any two spatially adjacent nodes $i$ and $j$, such an overlapping area is defined as 
$\mathcal{X}_{ij} = \mathcal{X}_i \cap \mathcal{X}_j$. 

In ad-hoc wireless networks, the decentralized topology 
is modeled as an undirected graph $\mathcal{G}_c = (\mathcal{A},\mathcal{E}_c)$, where an edge $(i,j)\in\mathcal{E}_c$ exists if and only if node $i$ and node $j$ share a non-empty physical overlap area $\mathcal{X}_{ij}\neq \emptyset$. 
The one-hop neighbor set of node $i$ is denoted by 
$N(i)=\{j \in \mathcal{A}:(i,j) \in \mathcal{E}_c\}$. 
To reduce communication overhead, neighboring nodes do not exchange information over the entire overlapping region. 
Instead, for node $i$, all its neighboring pairs  $(i,j), j\in N(i)$, forms a finite set of subsets, e.g., $Q$ subareas $\mathbf{b}_{ij,q}, q=1,\dots,Q$ are selected from node $i$'s overlapping areas with its neighbors to form a 
subset $\mathcal{B}_{ij} = \set{\mathbf{b}_{ij,q}}_{q=1}^{Q} \subset \mathcal{X}_{ij}, j\in N(i)$.
Such subset serves as anchors where neighboring nodes exchange 
low-dimensional information in decentralized collaborative training.


\textit{Remark-1:} In practical 
deployments, requiring 
distributed nodes to reach 
consensus at exactly matched  
spatial coordinates is 
unrealistic due to localization errors, mobility, and 
mismatched spatial resolutions. 
Therefore, we use small area-based monitoring anchors instead of coordinates. Such design not only improves robustness to spatial mismatch but also provides the flexibility to control the dimension of information exchanged among neighboring nodes.  

\subsection{Local Measurement Data}
During 
data collection, 
each 
node $i$ gathers a 
set of $N_i$ sparse PSD measurements 
within its local monitoring region: 
\begin{equation}\textstyle
	\mathcal{D}_i = \{(\boldsymbol{x}_{i,n}, \tilde{\mathbf{\Phi}}_{i,n})\}_{n=1}^{N_i}, \qquad \boldsymbol{x}_{i,n}\in \mathcal{X}_i,
	\label{eq:local_data}
\end{equation}
where $\tilde{\mathbf{\Phi}}_{i,n} = \mathbf{\Phi}(\boldsymbol{x}_{i,n}) + \nu_{i,n}$ represents the $n$-th ($n=1,\dots,N_i$)  PSD subject to noise 
$\nu_{i,n}$ observed by the mobile node $i$ at location $\boldsymbol{x}_{i,n}$. 
In addition to sparse PSD measurements, nodes may also access optional 
environmental features represented by 
a mapping $e(\cdot)\!:\! \mathcal{X} \!\rightarrow \!\mathbb{R}^{d_e}$, where the output of $e(\boldsymbol{x})$ 
encodes location-dependent 
physical descriptors, such as building heights, terrain elevation, or land-use information. 
Such environmental features 
further enhance RME if 
available. 

\subsection{Collaborative Training Objective}
The 
goal of decentralized RME in this work is to enable the distributed mobile edge nodes to collaboratively infer a 
consistent radio map 
$\hat{\mathbf{\Phi}}: \mathcal{X} \rightarrow \mathbb{R}$ together with 
a spatial variance map $\hat{\mathbf{v}}: \mathcal{X} \rightarrow \mathbb{R}_{+}$, where $\hat{\mathbf{v}}(\boldsymbol{x})$ characterizes the predictive uncertainty corresponding to the estimated $\hat{\mathbf{\Phi}}(\boldsymbol{x})$. 

Then, the decentralized learning aims to optimize the local neural network model parameters $\Theta_i$ for each node, such that collaborative training  minimizes the overall spatial reconstruction error and the uncertainty-aware negative log-likelihood over the target domain $\mathcal{X}$~\cite{kendall2017uncertainties}, which can be formulated as
\begin{equation}
	\min_{\set{\Theta_i}_{i=1}^K} \;
	\mathbb{E}_{(\mathbf{\Phi},e,\mathcal{D})} \{
		\sum_{\boldsymbol{x}\in\mathcal{X}_i}
		\mathcal{J}\big(\mathbf{\Phi}(\boldsymbol{x}), \hat{\mathbf{\Phi}}(\boldsymbol{x}), \hat{\mathbf{v}}(\boldsymbol{x})\big)
		\},
	\label{eq:global_objective}
\end{equation}
subject to the constraints of the communication graph $\mathcal{G}_c$ and the communication budget. The  construction of loss function $\mathcal{J}(\cdot)$ is specified next in Section~\ref{subsec:messaging}.

\begin{figure}
\centering
\includegraphics[width=0.7\linewidth]{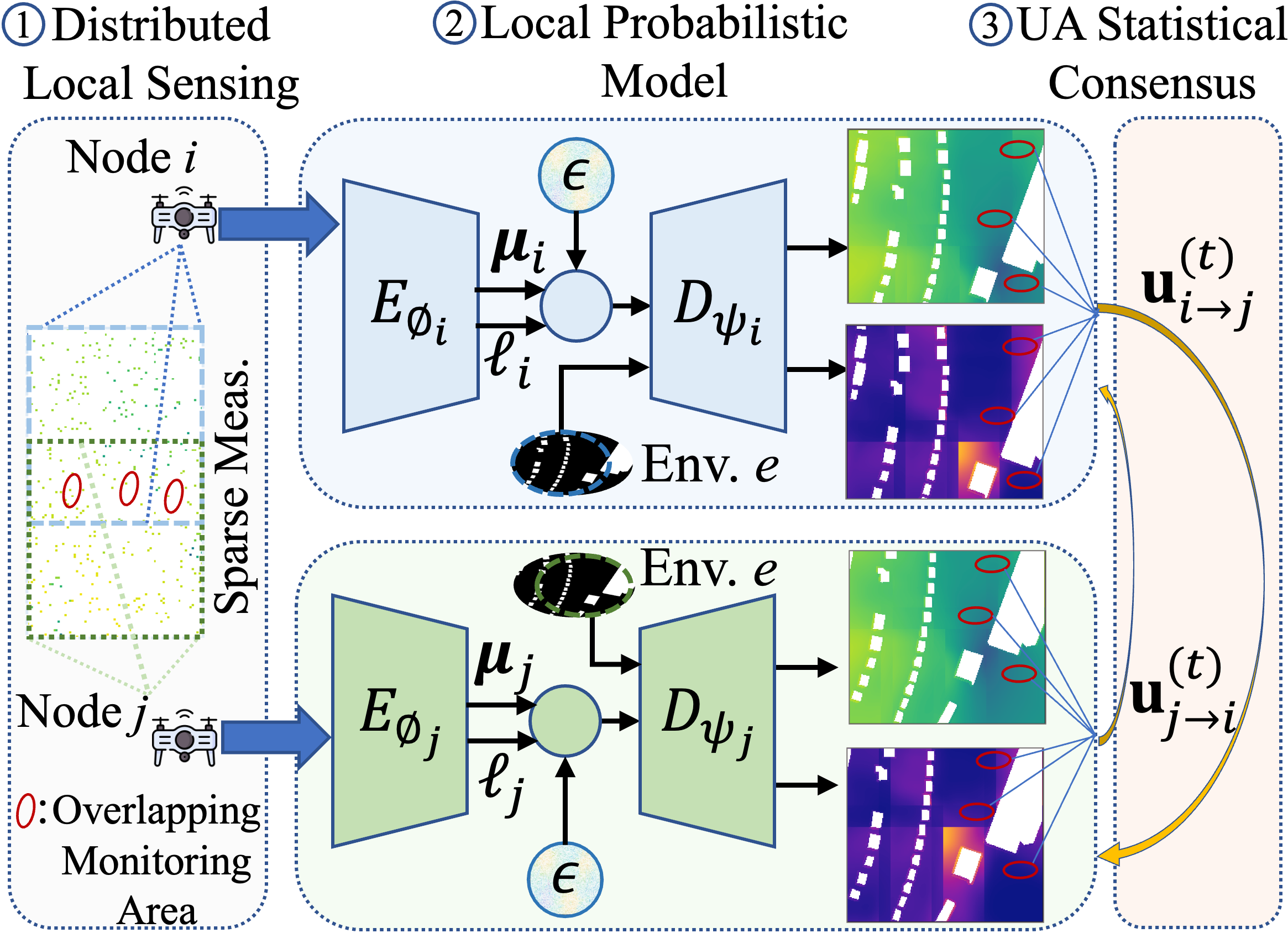}
\caption{Overall architecture of UA-D-RME framework.}
\label{fig:framework}
\vspace{-5mm}
\end{figure} 

\section{Uncertainty-Aware Distributed Radio Map Estimation}
\label{sec:method}
\subsection{Overall Architecture}
In the proposed UA-D-RME framework, 
to quantify the estimation uncertainty in unobserved regions, we employ the conditional variational autoencoder (cVAE)-type probabilistic neural network~\cite{kingma2013auto}.
In particular, each distributed mobile edge node maintains and trains its own local probabilistic cVAE model, whose goal is to map any spatial coordinates to the predicted PSD. 
To ensure estimation consistency across overlapping areas, UA-D-RME allows neighboring nodes to collaborate via an uncertainty-aware statistical consensus mechanism.
In collaborative training, neighboring nodes exchange only low-dimensional predictive statistics on demand, i.e., estimated PSD values and their uncertainty levels, evaluated at selected monitoring anchors in their overlapping areas. 
These exchanged statistics are then used to reach confidence-weighted consensus.
By incorporating these consensus mechanism into local training objectives, neighboring nodes iteratively align their local cVAE models, where RME estimates with lower uncertainty contribute more to the consensus.
Thus, UA-D-RME improves local reconstruction accuracy and neighborhood estimation consistency while maintaining low communication overhead.
The UA-D-RME architecture  is illustrated in Fig.~\ref{fig:framework}.

\subsection{Local Probabilistic Model}
\label{subsec:local_model}
Within the assigned monitoring region $\mathcal{X}_i$, each node $i$ trains a local cVAE model based 
on its sparse measurements $\mathcal{D}_i$. The goal is to infer both the expected PSD and the corresponding predictive uncertainty at any queried location $\boldsymbol{x}\in \mathcal{X}_i$. 

First, node $i$ transforms the raw measurement set $\mathcal{D}_i$ into a sequence-based input representation:
\begin{equation}
	X_i = \set{[E_f(\boldsymbol{x}_{i,n}),\, \tilde{\mathbf{\Phi}}_{i,n}] }_{n=1}^{N_i} \in \mathbb{R}^{N_i \times d_{\text{input}}},
	\label{eq:tokenized_input}
\end{equation}
where the transform operation $E_f(\cdot): \mathbb{R}^2 \rightarrow \mathbb{R}^{d_\text{Fourier}}$ applies a Fourier coordinate encoding, producing richer embeddings that facilitate modeling spatial irregularities~\cite{nan2026cgformer}. 

To learn the underlying distribution of $X_i$, the permutation-invariant encoder $E_{\phi_i}(\cdot)$ is applied via the DeepSets as~\cite{zaheer2017deep}: 
\begin{equation}
	(\boldsymbol{\mu}_i, \boldsymbol{\ell}_i) {=} 
    E_{\phi_i}(X_i){\triangleq}
    \rho_{\phi_i} \!( \frac{1}{N_i}\!\! \sum_{n=1}^{N_i}\! \varphi_{\phi_i} \!([E_f(\boldsymbol{x}_{i,n}), \tilde{\mathbf{\Phi}}_{i,n}], \, e(\boldsymbol{x}_{i,n}) ) ),
	\label{eq:posterior_init}
\end{equation}
where the two multilayer perceptron (MLP) networks $\varphi_{\phi_i}(\cdot)$ and $\rho_{\phi_i}(\cdot)$ 
constitute the encoder $E_{\phi_i}(\cdot)$. 
Within $E_{\phi_i}(\cdot)$,  $\rho_{\phi_i}(\cdot)$ maps the output of $\varphi_{\phi_i}(\cdot)$ to a $2d_z$-dimensional output, 
including the mean vector $\boldsymbol{\mu}_i \in \mathbb{R}^{d_z}$ and the variance vector $\boldsymbol{\ell}_i \in \mathbb{R}^{d_z}$, 
which serve as the distribution parameters of the posterior approximation applied in cVAE\cite{kingma2013auto}. 
Specifically, these parameters define a diagonal Gaussian posterior distribution over the latent 
variable $\mathbf{z}_i \in \mathbb{R}^{d_z}$:
\begin{equation}
	q_i(\mathbf{z}_i | X_i) = \mathcal{N}\big(\mathbf{z}_i;\boldsymbol{\mu}_i,\diag(\exp(\boldsymbol{\ell}_i))\big).
	\label{eq:local_posterior}
\end{equation}

It is worth noting that thanks to leveraging the statistical representation capability of the cVAE model customized for RME, it enables UA-D-RME to 
perform the quantification on estimation uncertainty in unobserved areas while maintaining end-to-end differentiability.
That is, the latent variable $\mathbf{z}_i$ is sampled using the reparameterization trick: $\mathbf{z}_i = \boldsymbol{\mu}_i + \boldsymbol{\epsilon} \odot \exp(\boldsymbol{\ell}_i/2)$, where $\boldsymbol{\epsilon} \sim \mathcal{N}(\mathbf{0}, \mathbf{I})$. Here, $\mathbf{z}_i$ acts as a stochastic propagation embedding conditioned on  local measurements $\mathcal{D}_i$.

Based on $\boldsymbol{x}$ and $\mathbf{z}_i$, 
the local radio map is then generated by a decoder $D_{\psi_i}(\cdot)$ at node $i$ as:

\begin{equation}
	\hat{\mathbf{\Phi}}_i(\boldsymbol{x}; \mathbf{z}_i) = D_{\psi_i}(\boldsymbol{x}, e(\boldsymbol{x}), \mathbf{z}_i), \qquad \boldsymbol{x} \in \mathcal{X}_i,
	\label{eq:local_field}
\end{equation}
where the local decoder $D_{\psi_i}(\cdot)$ is implemented as a MLP, and uses Gaussian error linear unit activations to ensure spatial differentiability of the estimated radio map. 

During training of UA-D-RME, the local probabilistic model draws samples of $\mathbf{z}_i$ from the latent distribution via \eqref{eq:local_posterior},
yielding the expected PSD values and their corresponding estimation variances at any geographical coordinate $\boldsymbol{x} \in \mathcal{X}_i$:
\begin{align}
	\hat{\mathbf{\Phi}}_i(\boldsymbol{x}) & = \mathbb{E}_{\mathbf{z}_i \sim q_i}\big[D_{\psi_i}(\boldsymbol{x}, e(\boldsymbol{x}), \mathbf{z}_i)\big],
	\label{eq:local_mean}                                                             \\
	\hat{\mathbf{v}}_i(\boldsymbol{x})              & = \mathrm{Var}_{\mathbf{z}_i \sim q_i}\big[D_{\psi_i}(\boldsymbol{x}, e(\boldsymbol{x}), \mathbf{z}_i)\big] + \sigma_{\min}^2,
	\label{eq:local_variance}
\end{align}
where the constant $\sigma_{\min}^2$ represents an irreducible 
noise floor, preventing numerical instability during the subsequent uncertainty-aware consensus processing.

\subsection{Uncertainty-Aware Statistical Consensus Training}
\label{subsec:messaging}
For decentralized collaborative training, to reach consensus between node $i$ and its neighbor $j$, 
at round $t$, node $i$ evaluates its current local model over $Q$ overlapping monitoring subareas $\mathcal{B}_{ij}{=}\{\mathbf{b}_{ij,1}, \dots, \mathbf{b}_{ij,Q}\}, j{\in} N(i)$.
This generates local statistical vectors for the overlapping subareas between node $i$ and its neighbors $j\in N(i)$: 
\begin{align}
	\hat{\mathbf{\Phi}}_{ij}^{(t)} & =
	\left[
		\hat{\mathbf{\Phi}}_{ij}^{(t)}(\mathbf{b}_{ij,1}), \ldots,
		\hat{\mathbf{\Phi}}_{ij}^{(t)}(\mathbf{b}_{ij,Q})
		\right]^\top,
	\label{eq:anchor_mean}           \\
	\mathbf{v}_{ij}^{(t)}            & =
	\left[
		\mathbf{v}_{ij}^{(t)}(\mathbf{b}_{ij,1}), \ldots,
		\mathbf{v}_{ij}^{(t)}(\mathbf{b}_{ij,Q})
		\right]^\top.
	\label{eq:anchor_var}
\end{align}

Then, the transmitted message from node $i$ to neighbors $j \in N(i)$ is the concatenation of these two kinds of statistics:
\begin{equation}
	\mathbf{u}_{i\rightarrow j}^{(t)} =
	\left[
		\hat{\mathbf{\Phi}}_{ij}^{(t)};\;
		\mathbf{v}_{ij}^{(t)}
		\right] \in \mathbb{R}^{2Q}.
	\label{eq:message}
\end{equation}
In fact, this statistics-based consensus design is 
communication-efficient. Because, instead of transmitting $N_i$ high-dimensional vectors in terms of model parameters or gradients, each node $i$ only exchanges $\mathcal{O}(Q)$ scalars with its  
neighbors. Furthermore, such a statistics-based message sharing mechanism remains physically interpretable to the proposed uncertainty-aware RME, 
which represents the local map estimates and their associated uncertainty in distributed RME, respectively.

Upon receiving such messages from its all neighboring nodes $j\in N(i)$,  node $i$ 
conducts a consensus process
through Gaussian conjugate prior update\cite{olfati2007distributed}:
\begin{subequations}\label{eq:consensus_targets}
	\begin{align}
		\bar{\mathbf{\Phi}}_{{ij}}^{(t)}
		 & =
		(
		\boldsymbol{\Lambda}_{ij}^{(t)} {+}\!\!\! \sum_{j\in N(i)}\!\!\! \boldsymbol{\Lambda}_{ji}^{(t)}
		)^{-1} 
		(
		\boldsymbol{\Lambda}_{ij}^{(t)}\hat{\mathbf{\Phi}}_{ij}^{(t)}
		{+}\!\!\! \sum_{j\in N(i)} \!\!\!\boldsymbol{\Lambda}_{ji}^{(t)}\hat{\mathbf{\Phi}}_{ji}^{(t)}
		),
		\label{eq:consensus_mean} \\
		\bar{\boldsymbol{\Lambda}}_{ij}^{(t)}
		 & =
		\boldsymbol{\Lambda}_{ij}^{(t)} + \!\!\!\sum_{j\in N(i)} \!\!\!\boldsymbol{\Lambda}_{ji}^{(t)},
		\label{eq:consensus_precision}
	\end{align}
\end{subequations}
where $\hat{\mathbf{\Phi}}_{ji}^{(t)}$ and $\mathbf{v}_{ji}^{(t)}$ represent the  estimated statistics of the neighbor node $j$ received at node $i$, and $\boldsymbol{\Lambda}_{ji}^{(t)} = \diag(1/\mathbf{v}_{ji}^{(t)})$ is the diagonal confidence 
matrix as the inverse of the received variances at node $i$ from node $j$, representing the reciprocal of estimation uncertainty.

Iteratively, node $i$ integrates such a consensus with its neighbors directly into its local training loss for the subsequent round $t+1$. The local objective 
at each node $i$ over its model parameters $\Theta_i = \{\phi_i, \psi_i\}$ becomes:
\begin{align}
	&\mathcal{J}_i^{(t)}(\Theta_i)
	  =
	\mathcal{L}_{\text{sup}}^{(i)} + \lambda_{\text{KL}} \text{KL}\big(q_i(\mathbf{z}_i)\,\|\, p(\mathbf{z}_i)\big)
	\nonumber + \\
	 &\lambda_{\text{comm}}
	\!\!\!\!\!\sum_{j\in \mathcal{N}(i)}
	\!\!\!\!\norm{
		\hat{\mathbf{\Phi}}_{ij} {-} \bar{\mathbf{\Phi}}_{ij}^{(t)}
	}_{\bar{\boldsymbol{\Lambda}}_{ij}^{(t)}}^2
    \!\! {+}
	\lambda_{\text{unc}}
	\!\!\!\!\!\sum_{j\in \mathcal{N}(i)}
	\!\!\!\!\norm{
		\log \mathbf{v}_{ij} {-} \log \bar{\mathbf{v}}_{ij}^{(t)}
	}_1,
	\label{eq:local_surrogate}
\end{align}
where $\bar{\mathbf{v}}_{ij}^{(t)} = \diag((\bar{\boldsymbol{\Lambda}}_{ij}^{(t)})^{-1})$, and the prior distribution $p(\mathbf{z}_i)$ is assumed to be $\mathcal{N}(\mathbf{0}, \mathbf{I})$\cite{kingma2013auto}. 
The local supervision loss $\mathcal{L}_{\text{sup}}^{(i)}$  enforces alignment of estimated radio map values with the ground-truth data in local monitoring region of interest $\mathcal{X}_i$, 
and the associated variance $\mathbf{v}_i(\boldsymbol{x})$ is defined as:

\begin{equation}
	\mathcal{L}_{\text{sup}}^{(i)}
	{=}
	\!\!\sum_{\boldsymbol{x}\in \mathcal{X}_i}
	\!\!(
		\frac{1}{2\hat{\mathbf{v}}_i(\boldsymbol{x})}\norm{\mathbf{\Phi}_i(\boldsymbol{x}){-}\hat{\mathbf{\Phi}}_i(\boldsymbol{x})}^2
		\!\!{+}
		\frac{1}{2}\log \hat{\mathbf{v}}_i(\boldsymbol{x})
		).
	\label{eq:local_nll}
\end{equation}

This cVAE-enabled, statistics-based consensus scheme ensures that each node optimizes its local probabilistic model to  estimate the radio map and  uncertainty from its own sparse data, while simultaneously penalizing inconsistency on local estimation over overlapping areas 
with its neighbors. 
For neighbors who generate predictions with higher  confidence (lower variance) contribute more in reaching neighborhood consensus on collaborative training.

\section{Theoretical Analysis}
\label{sec:theory}
We provide a rigorous theoretical analysis of the proposed UA-D-RME method. We investigate two fundamental properties: the convergence of the proposed uncertainty-aware consensus mechanism during training, 
and the overall spatial reconstructability of the distributed architecture. 
Due to the page limit, the detailed proofs of the following theorems can be found in the supplementary of this work\footnote{The proofs are available at {\url{https://github.com/DongYang26/UA-D-RME/blob/main/Supplementary.pdf}}.}.

\subsection{Convergence of Uncertainty-Aware Statistics Consensus}
A 
concern in decentralized learning without centralized coordination of global model information exchange is whether the local models can reach an 
agreement on the overlapping monitoring areas. 
We analyze the convergence of the local estimates over 
$\mathcal{B}_{ij}$ driven by the local surrogate loss in \eqref{eq:local_surrogate}. 

\begin{theorem}[Convergence of Boundary Mismatches]
\label{thm:convergence}
Under the condition that the local neural decoders $D_{\psi_i}$ are sufficiently over-parameterized and the communication graph $\mathcal{G}_c$ is connected, the boundary estimates updated via gradient descent on $\mathcal{J}_i^{(t)}(\Theta_i)$ follow an 
uncertainty-aware Laplacian consensus optimization.  
For a properly chosen learning rate $\eta < \frac{2}{\lambda_{\max}(\mathbf{L}_{\boldsymbol{\Lambda}})}$, the sequence of inconsistency at overlapping monitoring areas converges exponentially to a bounded neighborhood defined by the irreducible local supervision gradients:
\begin{equation}\textstyle
    \lim_{t \to \infty} \mathbb{E} \{ \sum_{(i,j)\in \mathcal{E}_c} \!\!\norm{\hat{\mathbf{\Phi}}_{ij}^{(t)} {-} \hat{\mathbf{\Phi}}_{ji}^{(t)}}^2_{\bar{\boldsymbol{\Lambda}}_{ij}^{(t)}} \} \!\leq\! \mathcal{O}\!\!\left( \frac{\sigma_{\mathrm{local}}^2}{\lambda_2(\mathbf{L}_{\boldsymbol{\Lambda}})} \right),
\end{equation}
where $\mathbf{L}_{\boldsymbol{\Lambda}}$ is the uncertainty-aware 
Laplacian matrix of the communication graph, the second-smallest eigenvalue  $\lambda_2(\mathbf{L}_{\boldsymbol{\Lambda}}) > 0$ of  $\mathbf{L}_{\boldsymbol{\Lambda}}$ determines 
its algebraic connectivity, and $\sigma_{\mathrm{local}}^2$ denotes the bounded variance of the local supervision loss gradients.
\end{theorem}

Theorem~\ref{thm:convergence} 
indicates that the proposed UA-D-RME framework 
guarantees the convergence behavior. 
The convergence rate and the final consensus gap heavily depend on $\lambda_2(\mathbf{L}_{\boldsymbol{\Lambda}})$, implying that neighbors with higher prediction confidence (i.e., larger precision weights of $\boldsymbol{\Lambda}$) accelerate the consensus and enforce tighter boundary consistency.

\subsection{Spatial Reconstructability and Generalization Error Bound}
Then, we evaluate the global generalization capability of the decentralized cVAE-based architecture. A key challenge is ensuring that the local estimators, trained on non-IID sparse local datasets $\mathcal{D}_i$, can collectively bound the expected global loss in~\eqref{eq:global_objective}.

Let $\mathcal{R}(\hat{\mathbf{\Phi}}) = \mathbb{E}_{\boldsymbol{x} \sim \mathcal{X}}[ |\mathbf{\Phi}(\boldsymbol{x}) - \hat{\mathbf{\Phi}}(\boldsymbol{x})| ]$ denote the expected global generalization error. 
Let $\mathfrak{R}_{N_i}(D_{\psi_i})$ represent the empirical Rademacher complexity of the local decoder hypothesis space with respect to the $N_i$ samples in $\mathcal{D}_i$.

\begin{theorem}
[Distributed Generalization Error Bound]
\label{thm:generalization}
With probability at least $1-\delta$, the expected global spatial reconstruction error is upper-bounded by:
\begin{align}\textstyle
    \mathcal{R}(\hat{\mathbf{\Phi}}) \leq \sum_{i=1}^K p_i [ & \hat{\mathcal{R}}_{\mathcal{D}_i}(\hat{\mathbf{\Phi}}_i) + 2 \mathfrak{R}_{N_i}(D_{\psi_i}) \nonumber \\
     + 3M_{\Phi}&\sqrt{{\ln(2K/\delta)}/{2N_i}} + \mathcal{O}(\Delta_{\mathrm{b}}) ],
\end{align}
where $p_i = \int_{\mathcal{X}_i} \alpha_i(\boldsymbol{x}) p(\boldsymbol{x}) d\boldsymbol{x}$ is the effective spatial contribution of node $i$, $\alpha_i(\boldsymbol{x})$ is node $i$'s contribution to the global estimate at $\boldsymbol{x}$, $\hat{\mathcal{R}}_{\mathcal{D}_i}(\hat{\mathbf{\Phi}}_i)$ is the local empirical reconstruction risk, $M_{\Phi}$ is the supremum bound of the PSD, and $\mathcal{O}(\Delta_{\mathrm{b}})$ represents the residual boundary mismatch established in Theorem~\ref{thm:convergence}.
\end{theorem}

Theorem~\ref{thm:generalization} 
demonstrates that the reconstructability of UA-D-RME is independent  
from the requirement on centralized training. As long as each node collects a modest number of samples $N_i$ to suppress local Rademacher complexity, and the statistical consensus minimizes $\Delta_{\mathrm{b}}$, the 
estimated radio map is 
guaranteed to converge to the true PSD distribution.

\section{Numerical Results}
\label{sec:experiments}
\subsection{Dataset}
We evaluate UA-D-RME on urban radio maps generated from RadioMapSeer~\cite{levie2021radiounet} under a decentralized single-band setting. The dataset captures heterogeneous urban propagation environments from six major cities, including Ankara, Berlin, Glasgow, Ljubljana, London, and Tel Aviv. The dataset covers 700 spatial scenes, with each scene providing 80 distinct radio maps. In the simulation setup, the transmitters operate at a 5.9 GHz carrier frequency with a radiated power of 23 dBm.

\subsection{Simulation Setups}
To test spatial generalization, the 700 distinct scenes are split into 600 training, 50 validation, and 50 test scenes. Within each scene, the target is rasterized into a 256$\times$256 spatial grid and partitioned into a 2$\times$4 topology of overlapping sensing regions, monitored by $K=8$ distributed mobile edge nodes. Adjacent regions maintain a 15\% coverage overlap, providing $Q=16$ discrete monitoring areas per boundary as consensus anchors. To emulate practical unbalanced network deployments, nodes acquire sparse measurements strictly within their assigned areas. For each map, only 0.5\% of the measurement data is non-uniformly sampled across all nodes, enforcing explicitly heterogeneous local observations.

To ensure a fair comparison, all baselines employ the identical probabilistic neural backbone as UA-D-RME. Specifically, we compare against standalone learning-based RME (S-RME), an isolated local modeling approach with strictly zero inter-node collaboration, and centralized learning based RME (C-RME), an idealized architecture with unconstrained access to the globally aggregated measurements that serves as the 
performance upper bound.

To comprehensively evaluate both the 
RME reconstruction fidelity and the quality of the spatial uncertainty estimation, we employ four metrics: Root Mean Square Error (RMSE), Negative Log-Likelihood (NLL), Prediction Interval Coverage Probability (PICP), and Mean Prediction Interval Width (MPIW). RMSE quantifies the absolute accuracy of the reconstructed radio map, while NLL evaluates the overall probabilistic modeling quality. Furthermore, PICP and MPIW are utilized jointly to assess whether the predicted spatial uncertainty is both well-calibrated and sufficiently sharp. Specifically, PICP represents the empirical coverage probability corresponding to the 90\% prediction confidence. 
A well-calibrated probabilistic model is expected to achieve a PICP 
approximating 0.90 while simultaneously maintaining a narrow MPIW, thereby providing highly informative confidence bounds for downstream applications.

\begin{figure}
\centering
\includegraphics[width=0.82\linewidth]{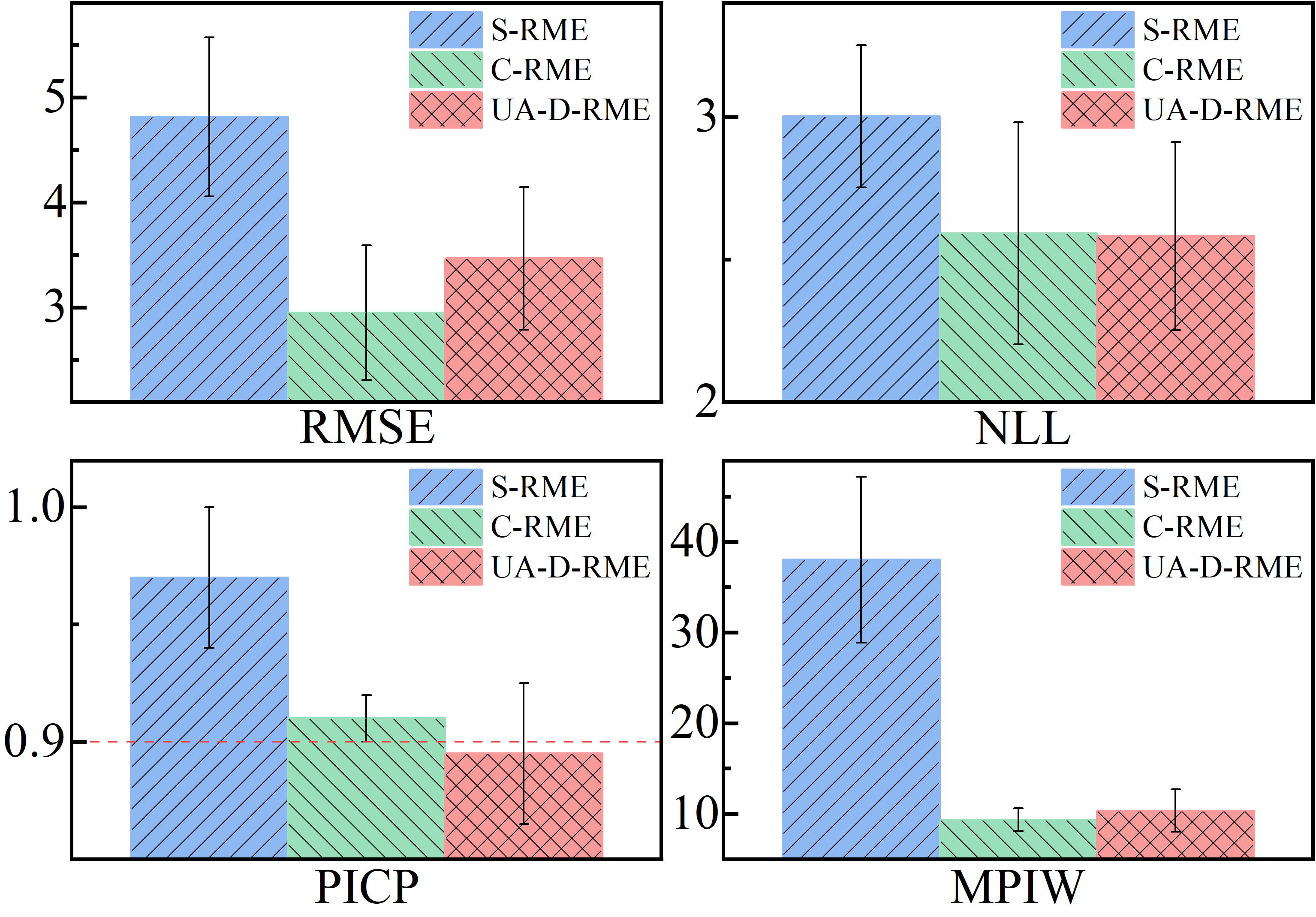}
\caption{Estimation performance in different methods.}
\label{fig:Evalu}
\end{figure} 

\subsection{Evaluation and Comparison}
As shown in Fig.~\ref{fig:Evalu}, 
S-RME suffers from data sparsity, yielding the worst RMSE and uninformative confidence intervals (i.e., an excessively large MPIW despite a high PICP), which is also reflected by its poor NLL. As the centralized upper bound, C-RME achieves the best overall accuracy and sharpest intervals, but necessitates prohibitive raw data aggregation.
In contrast, the proposed UA-D-RME significantly narrows the performance gap to C-RME across all metrics. Notably, it achieves near-optimal NLL and tight confidence bounds (substantially reducing MPIW compared to S-RME), while satisfying the target PICP threshold. 
Meanwhile,  
UA-D-RME consumes 
low overhead during collaboration, 
by exchanging only low-dimensional statistics on overlapping areas.  

\begin{figure}
\centering
\includegraphics[width=1\linewidth]{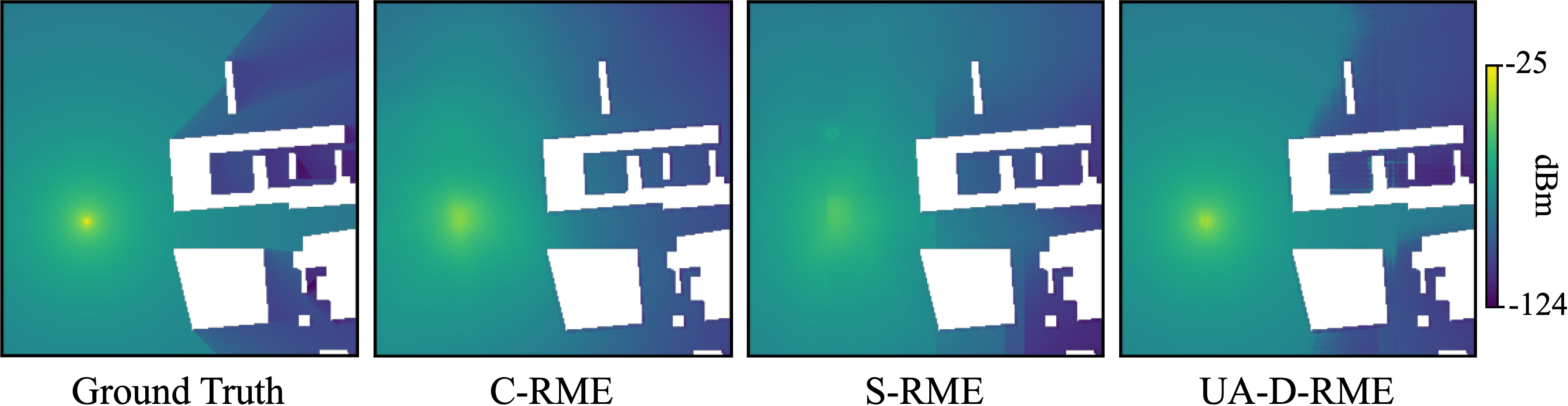}
\caption{Visualization of the reconstructed global radio 
maps.}
\vspace{-2mm}
\label{fig:Visuali}
\end{figure} 
\begin{figure}
\centering
\includegraphics[width=1\linewidth]{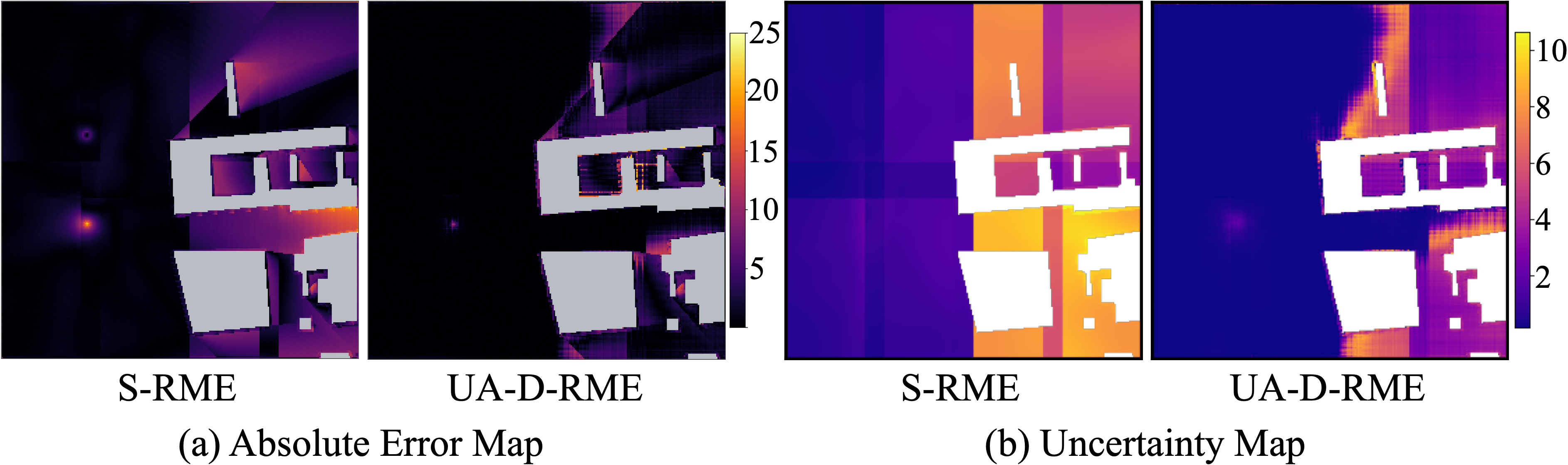}
\caption{Visualization of spatial distributions of absolute prediction errors and estimated uncertainties.}
\vspace{-2mm}
\label{fig:Addit_maps}
\end{figure}

As shown in Fig.~\ref{fig:Visuali} and Fig.~\ref{fig:Addit_maps} (a), visualizations corroborate these advantages and quantitative findings. 
The proposed UA-D-RME method successfully eliminates the severe spatial discontinuities that plague the S-RME results, producing seamless transitions across the overlapping coverage regions. More importantly, as shown in Fig.~\ref{fig:Addit_maps} (b), the 
uncertainty maps demonstrate profound physical significance: high estimation confidence is predominantly concentrated around densely sensed areas and shared monitoring areas with strong multi-node consensus. In contrast, elevated estimation uncertainty (larger variances) 
emerges near severe propagation blockages, and regions with sparse observations. This well-calibrated spatial uncertainty reflects the reliability of the reconstructed radio environment, providing crucial information for downstream network planning and resource allocation.

\section{Conclusion}
\label{sec:conclusion}
This paper proposed an uncertainty-aware distributed radio map estimation framework (UA-D-RME) for decentralized 
wireless networks.
Unlike centralized RME and conventional federated learning approaches, UA-D-RME avoids raw measurement collection and high-dimensional model-update exchange.
By incorporating estimation uncertainty into statistical consensus, UA-D-RME allows distributed nodes to align their local probabilistic models in a confidence-weighted manner, improving both reconstruction accuracy and uncertainty calibration under communication constraints.
Theoretical analysis demonstrates the convergence behavior of the proposed consensus mechanism and characterizes the distributed generalization performance. 
Numerical results on urban radio map datasets verify that UA-D-RME achieves reconstruction performance close to the centralized benchmark while reducing communication overhead and without exchanging raw measurements. 

\bibliographystyle{IEEEtran}
\bibliography{bib}
\end{document}